\documentclass[prb,reprint,showkeys,showpacs]{revtex4-1}
\usepackage{graphicx}
\newcommand{\mbra}[1]{\mathinner{\langle{#1}|}}
\newcommand{\mket}[1]{\mathinner{|{#1}\rangle}}

\newcommand{\ket}[1]{$\mket{#1}$}
\newcommand{\identity}[0]{\hat \mathcal I}
\newcommand{\sdownarrow}[0]{\downarrow}
\newcommand{\suparrow}[0]{\uparrow}
\newcommand{\mup}[0]{${\cal M}_{\uparrow}$}
\newcommand{\mdown}[0]{${\cal M}_{\downarrow}$}

\begin{document}



\title{Tunneling spectroscopy using a probe qubit}
\author{A. J. Berkley}\email{ajb@dwavesys.com}
\author{A. J. Przybysz}
\author{T. Lanting}
\author{R. Harris}
\author{N. Dickson}
\altaffiliation[Current address: ]{Side Effects Software Inc. 1401-123 Front St. West, Toronto ON Canada M5J 2M2}
\author{F. Altomare}
\author{M. H. Amin}
\author{P. Bunyk}
\author{C. Enderud}
\author{E. Hoskinson}
\author{M. W. Johnson}
\author{E. Ladizinsky}
\author{R. Neufeld}
\author{C. Rich}
\author{A. Yu. Smirnov}
\author{E. Tolkacheva}
\author{S. Uchaikin}
\author{A. B. Wilson}
\affiliation{D-Wave Systems Inc. 100-4401 Still Creek Dr., Burnaby BC Canada V5C 6G9}
\begin{abstract}
We describe a quantum tunneling spectroscopy technique that requires only low bandwidth control.  The method involves coupling a probe qubit to the system under study to create a localized probe state.  The energy of the probe state is then scanned with respect to the unperturbed energy levels of the probed system.  Incoherent tunneling transitions that flip the state of the probe qubit occur when the energy bias of the probe is close to an eigenenergy of the probed system.  Monitoring these transitions allows the reconstruction of the probed system eigenspectrum.  We demonstrate this method on an rf SQUID flux qubit.
\end{abstract}
\maketitle


Recent technological advances have allowed the construction of mesoscale systems of individual quantum elements, including hundreds of trapped ions\cite{Britton2012}, 14 entangled ions\cite{PhysRevLett.106.130506}, nanomagnetic systems assembled out of magnetic atoms on metallic surfaces\cite{Khajetoorians2012}, ultracold $^{87}$Rb atoms in optical lattices\cite{Weitenberg2011}, and arrays of superconducting devices\cite{Johnson2011, Mariantoni07102011}.  While the study of small numbers of atoms or devices often involves direct manipulation and full state tomography, these techniques become impractical in the mesoscale regime. As a result, there is a need for tools that are applicable when one has a mesoscale system with limited control over its individual elements.

Tunneling spectroscopy is a powerful tool for studying condensed
matter systems.  It has been used to push the limits of our
understanding of many-body physics, as in recent studies of two
dimensional electron systems in high magnetic field using time domain
capacitance spectroscopy\cite{Dial2010} and scanning tunneling
spectroscopy \cite{Song2010}.  Tunneling spectroscopy can also be used
to directly validate numerical or analytical models of complex
systems, such as the single particle states of CdSe quantum dots
\cite{Bakkers2001} or the electronic wavefunctions of carbon
nanotubes\cite{Lemay2001}.  Motivated by the ability of tunneling
spectroscopy to probe the quantum behaviour of mesoscale systems, we
have developed an analogous method that is applicable when one has
limited control over a large system.  The large
system is probed using a dedicated probe qubit with its own
readout and low bandwidth control of its Hamiltonian.  We have termed
this new technique qubit tunneling spectroscopy (QTS).

A related method has been proposed in
Ref.~\onlinecite{PhysRevA.85.062304} where the probe qubit must be
perturbatively coupled to the system under study.  In QTS, the
requirement for this weak coupling has been removed through the use of
a compensation bias (as explained below).  Further, the algorithm of
Ref.~\onlinecite{PhysRevA.85.062304} is designed to operate on a gate
model quantum computer while we demonstrate QTS on a system with much
more limited control.

QTS requires a probe qubit $P$ that can be described by a generic
two-level system Hamiltonian:
\begin{equation}\label{eqn:Hisingspin}
  \hat H_{P} = -\frac{1}{2}\Delta_P \hat \sigma_{x,P} - \frac{1}{2}\epsilon_P \hat\sigma_{z,P},
\end{equation}
where $\hat\sigma_{x,P}$ and $\hat\sigma_{z,P}$ are Pauli matrices
operating on $P$, and both parameters $\epsilon_P$ and $\Delta_P$ should be
controllable.  The eigenstates of $\hat\sigma_{z,P}$ with eigenvalues
+1 and -1 are $\mket{\suparrow}_P$ and $\mket{\sdownarrow}_P$,
respectively.  The $\hat\sigma_{z,P}$ eigenstates should be
distinguishable by a readout mechanism.  Let there be a system $S$,
governed by some Hamiltonian $\hat H_S$, that one would like to study.

To perform QTS we require a coupling between the probe qubit and a parameter of the system (described by an operator $\hat C$) as well as a controllable compensation bias $\epsilon_{\text{comp}}$ coupled to that same parameter.  In this case, the system plus probe Hamiltonian can be expressed as:
\begin{equation}\label{eqn:coupledsystem}
  \hat H_{S+P} = \hat H_S + \hat H_P + J \hat \sigma_{z,P} \hat C + \frac{1}{2} \epsilon_{\text{comp}} \hat C ,
\end{equation}
with $J$ the strength of the probe qubit-system interaction.

For general $\epsilon_{\text{comp}}$ the eigenstates of $S+P$ are not
representative of those of $S$.  However, in the special case
$\epsilon_{\text{comp}} = -2J$, the spectrum of $\hat H_{S+P}$ splits
into two qualitatively different manifolds, \mup\ and \mdown, wherein
the probe qubit $P$ is in state $\mket{\suparrow}_P$ and
$\mket{\sdownarrow}_P$, respectively. $H_{S+P}$ can then be rewritten
as:
\begin{eqnarray}\label{eqn:manifold}
  \hat H_{S+P} = & \left(\epsilon_P \identity_S - 2 J \hat C + \hat H_S\right)\otimes\mket{\sdownarrow}_P\mbra{\downarrow}_P \\ \nonumber
                   & + \hat H_S\otimes \mket{\suparrow}_P\mbra{\uparrow}_P \\ \nonumber
                   & - \frac{\Delta_P}{2}
                     \identity_S
                     \otimes\Big(\mket{\sdownarrow}_P\mbra{\uparrow}_P + \mket{\suparrow}_P\mbra{\downarrow}_P\Big)
\end{eqnarray}
where $\identity_S$ is the identity operator on system $S$.  In the
case where the third line of Eqn.~\ref{eqn:manifold} is perturbatively
small, the first line is the Hamiltonian of \mdown\ and the second
line that of \mup.  Thus, the energy spectrum of \mup\ is identical to
that of $\hat H_S$.  Further, the first term of the first line of
Eqn.~\ref{eqn:manifold} shows that the energy of all states in \mdown\
can be shifted with respect to those of \mup\ by adjusting the probe
energy bias $\epsilon_P$.  For small enough $\Delta_P$, the third line of
Eqn.~\ref{eqn:manifold} gives rise to incoherent inter-manifold
tunneling\cite{PhysRevLett.100.197001} between any state
$\mket{k'_\downarrow}\mket{\downarrow}_P$ of \mdown\ and any state
$\mket{k}\mket{\uparrow}_P$ of \mup\ with a rate proportional to
$\mathinner{|\Delta_P\mathinner{\langle k'_\downarrow|k\rangle}|^2}$.

The QTS method begins by initializing the system into the lowest
energy state of \mdown.  The tunneling rate between manifolds peaks
when an eigenstate of \mup\ is brought into resonance with the initial
system state in \mdown\ by adjusting $\epsilon_P$.  This resonant
tunneling transition between manifolds flips the state of the probe
qubit, which can be easily detected.  Thus, to perform QTS, one
measures the initial transition rate
$\Gamma\equiv\left|dP/dt\right|_{t=0}$, where $P$ is the probability
of observing the probe qubit in its initial state, as a function of
$\epsilon_P$, the probe energy bias.  Scanning $\epsilon_P$ and
locating peaks in $\Gamma$ allows one to map out the eigenspectrum of
\mup\ which is identical to that of $\hat H_S$ if the compensation
bias $\epsilon_{\text{comp}}$ is set to $-2 J$.  Note that errors in
this compensation bias will skew the energy spectrum of the probed
system.  The errors in the extracted energy spacings of the spectrum
are bounded by the compensation bias error.

\begin{figure}
\includegraphics{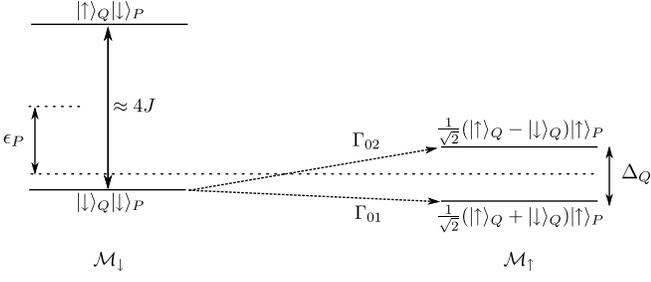}
\caption{\label{fig:manifoldenergies} Energy level diagram for the $Q+P$ qubit system described by the Hamiltonian in Eqn.~\ref{eqn:HQ+P} at $\epsilon_Q = 0$ and $\epsilon_{\text{comp}} = - 2 J$. The system separates into two distinct manifolds, \mdown\ and \mup, that differ in the orientation of the probe qubit $P$. Spinor notation indicates state of qubits $Q$ and $P$ on the left and right, respectively. Allowing for weak tunneling in qubit $P$ facilitates incoherent tunneling processes $\Gamma_{01}$ and $\Gamma_{02}$ between the manifolds.  QTS proceeds by initializing the system in the lowest energy state of \mdown and finding values of $\epsilon_P$ for which transition rate $\Gamma$ peaks due to resonant processes $\Gamma_{01}$ and $\Gamma_{02}$.}
\end{figure}

To experimentally demonstrate QTS we take a target system comprising a
single qubit $Q$, governed by a Hamiltonian $\hat H_Q$
(Eqn.~\ref{eqn:Hisingspin} with $P\rightarrow Q$).  We couple the
probe qubit $P$ with strength $J$ through a mutual $\hat \sigma_z$
interaction to qubit $Q$.  In this case, $\hat H_S \rightarrow \hat
H_Q$ and $\hat C \rightarrow \hat \sigma_{z,Q}$ in
Eqn.~\ref{eqn:coupledsystem}, yielding
\begin{equation}\label{eqn:HQ+P}
\hat H_{Q+P} = \hat H_{Q}  + \hat H_{P} + J  \hat \sigma_{z,P} \hat \sigma_{z,Q}  + \frac{1}{2} \epsilon_{comp} \hat \sigma_{z,Q}.
\end{equation}
$\Delta_P$ is chosen to be small compared with all other terms so the
eigenstates of the probe are to good approximation
$\mket{\suparrow}_P$ and $\mket{\sdownarrow}_P$.  Setting
$\epsilon_{comp} = -2 J$ then yields a Hamiltonian of the form given
in Eqn.~\ref{eqn:manifold}.  For the particular case $\epsilon_Q = 0$, we show how the  theoretical energy spectrum of the coupled two qubit system splits into two manifolds in Fig.~\ref{fig:manifoldenergies}.  Note that the
eigenstates of \mup\ shown therein are superpositions of the
$\mket{\suparrow}_Q$ and $\mket{\sdownarrow}_Q$ states.  Changing
$\epsilon_P$ allows the lowest state in \mdown,
$\mket{\sdownarrow}_Q\mket{\sdownarrow}_P$, to be brought into
resonance with the states in \mup.  The system plus probe can tunnel
from \mdown\ to \mup\ through the incoherent processes labelled as
$\Gamma_{01}$ and $\Gamma_{02}$.

The two qubits, probe $P$ and target $Q$, used in the experiment were
rf SQUID flux qubits on a D-Wave quantum annealing processor
thermalized to a temperature $T=12\,$mK.  A description of a chip
similar to that used in this study can be found in
Ref.~\onlinecite{eightqunitcell}.  The low energy rf SQUID flux qubit
Hamiltonian \cite{robustqubit} has a direct mapping onto
Eqn.~\ref{eqn:Hisingspin}:
\begin{equation}\label{eqn:1qhamiltonian}
\hat H_{Q} = -\frac{1}{2}\Delta_{Q}(\Phi^{ccjj}_Q) \hat\sigma_{x,Q} - \Phi^x_Q \left|I^p_Q(\Phi^{ccjj}_Q)\right| \hat\sigma_{z,Q} 
\end{equation}
where we have performed the substitution $\Delta_Q \rightarrow
\Delta_Q(\Phi^{ccjj}_Q)$ and $\epsilon_Q \rightarrow 2
\left|I^p_Q(\Phi^{ccjj}_Q)\right|\Phi^x_Q$, with $\Phi^{ccjj}_Q$ and
$\Phi^x_Q$ being externally controlled flux biases and
$\left|I^p_Q(\Phi^{ccjj}_Q)\right|$ being the magnitude of the qubit
persistent current.  Note that both $\Delta_Q$ and $\epsilon_Q$
(through $\left|I^p_Q\right|$) are functions of $\Phi^{ccjj}_Q$.  The
functional forms of these dependencies are determined by the physical
parameters of the rf SQUID, as described in detail in
Ref.~\onlinecite{robustqubit}.  If one considers the qubit $Q$ as an
Ising spin, then $\Delta_Q$ corresponds to a transverse magnetic
field, $I^p_Q$ is the magnitude of the spin, and $\Phi^x_Q$ is an
applied longitudinal magnetic field.  The physical Hamiltonian for the
probe qubit $\hat H_{P}$ is found by replacing $Q$ by $P$ in
Eqn.~\ref{eqn:1qhamiltonian}.  The probe qubit had a persistent
current $\big|I^p_P\big|=1.0\,\mu$A and $\Delta_P/h \sim 1$ MHz.  The
small $\Delta_P$ was chosen so that the transition rate $\Gamma$ of
the probe qubit was contained within the dc to 3 MHz bandwidth of our
slow control lines and to satisfy the incoherent inter-manifold
tunneling condition.  An on-chip tunable coupler\cite{eightqunitcell}
between the two qubits was programmed to attain an interqubit mutual
inductance $M=2.0\,$pH.  The resulting form for $J$ in
Eqn.~\ref{eqn:HQ+P} is $J = M \big|I^p_Q\big|\big|I^p_P\big|$.  With
these parameters, the spectral gap in the \mdown\ manifold, as
depicted in Fig.~\ref{fig:manifoldenergies}, satisfied
$4J=4M\big|I^p_Q\big|\big|I^p_P\big|\gg k_BT$ over the range of
$\big|I^p_Q\big|$ encountered in these experiments.  Consequently,
there was negligible thermal activation out of the initial state
\ket{\downarrow}$_Q$\ket{\downarrow}$_P$ to higher levels within
\mdown.  With these parameters, the compensation bias is explicitly
$\frac{1}{2}\epsilon_{comp}\hat\sigma_{z,Q} = -J \hat \sigma_{z,Q} =
-M \big| I^p_Q \big| \big| I^p_P \big| \hat \sigma_{z,Q}$ and is
applied by adding an offset $M \big| I^p_P \big| \sim 1$ m$\Phi_0$ to
the flux bias $\Phi^x_Q$ of qubit $Q$.  This compensation bias
requires only careful calibration of probe parameters.
 
The experimental method for initialization and readout is the same as
the two-qubit cotunneling technique described in
Ref.~\onlinecite{2qmrt}.  The experiments described herein differed
from the cotunneling experiment in three regards: First, in QTS one
intentionally sets $\Delta_P\ll\Delta_Q$, thus exploring an extreme
limit of the mismatched tunneling energy configuration described in
Ref.~\onlinecite{2qmrt}.  Second, in QTS we use relatively large
offset biases $\epsilon_{comp}/\big|I^p_Q\big| = 1 $m$\Phi_0$ in order
to satisfy the compensation condition embodied in
Eqn.~\ref{eqn:manifold}.  Third, whereas the dynamics studied in
Ref.~\onlinecite{2qmrt} involved incoherent tunneling of the pair of
qubits between localized initial and final spin states, in QTS the
final state can place qubit $Q$ in a delocalized (superposition)
state, as depicted in Fig.~\ref{fig:manifoldenergies}.

\begin{figure}
\includegraphics{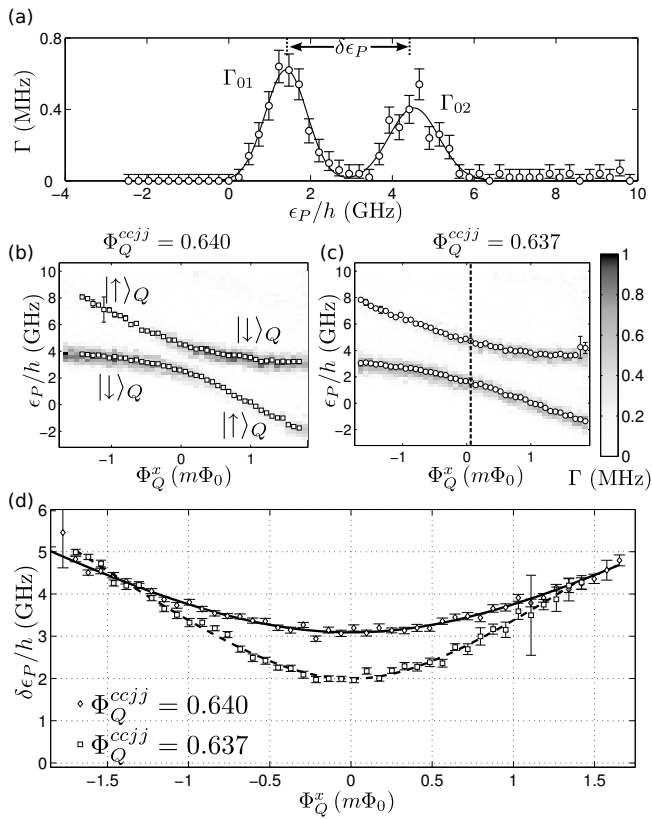}
\caption{\label{fig:sample-1q-spectrum} (a) QTS
  tunneling rate data versus $\epsilon_p$ for
  $\Phi^x_Q=0.1\,$m$\Phi_0$ and $\Phi^{ccjj}_Q = 0.637$.  Peaks in
  $\Gamma$ are readily associated with the processes $\Gamma_{01}$ and
  $\Gamma_{02}$ denoted in Fig.~\ref{fig:manifoldenergies}.  The
  centers of the peaks are found by fitting to a sum of two Gaussian
  peaks.  (b) Qubit $Q$ energy spectra obtained by QTS at control bias
  values: $\Phi^{ccjj}_{Q}/\Phi_0 = 0.640$; (c)
  $\Phi^{ccjj}_{Q}/\Phi_0=0.637$.  In both plots, the ordinate is the
  probe bias energy $\epsilon_P = 2 \big|I^p_P\big| \Phi^x_P$ and the
  abscissa is the flux bias $\Phi^x_Q \propto \epsilon_Q$ applied to
  qubit $Q$.  The grayscale indicates the transition rate $\Gamma$ in the
  $(\Phi^x_Q,\epsilon_P)$ plane.  White circles denote the centers of
  peaks in $\Gamma$ found by the Gaussian fits.  Avoided crossings
  between two localized states, explicitly labelled in (b) as
  \ket{\uparrow}$_Q$ and \ket{\downarrow}$_Q$, are visible.  (d) The
  difference in probe energy $\delta\epsilon_P$ between the two peak
  centers as a function of $\Phi^x_q$ for the datasets in (a) and (b).
  Results have been fit to the dispersion of the Hamiltonian in
  Eqn.~\ref{eqn:1qhamiltonian} using $\Delta_Q$ and
  $\left|I^p_Q\right|$ as free parameters.}
\end{figure}

A scan of the initial transition rate $\Gamma$ versus $\epsilon_P$ at
$\Phi^x_Q \sim 0$ and $\Phi^{ccjj}_Q= 0.637$ is shown in
Fig.~\ref{fig:sample-1q-spectrum}(a).  The data clearly show two
distinct peaks.  These peaks are readily identified as the processes
$\Gamma_{01}$ and $\Gamma_{02}$ indicated in
Fig.~\ref{fig:manifoldenergies}.  We fit such scans to a model
composed of a pair of Gaussian peaks in order to locate the peak
centers.  Example maps of the initial transition rate $\Gamma$ versus
$\epsilon_P$ for a range of qubit $Q$ flux biases around $\Phi^x_Q=0$
are shown in Figs.~\ref{fig:sample-1q-spectrum}(b) and (c) for two
values of the control flux bias $\Phi^{ccjj}_Q$ (and therefore
$\Delta_Q$).  For clarity the centers of the Gaussian peaks have been
indicated by white circles.  The peak positions reveal the avoided
crossing between localized spin states $\mket{\suparrow}_Q$ and
$\mket{\sdownarrow}_Q$.  In Fig.~\ref{fig:sample-1q-spectrum}(d), we
summarize the difference in probe energy $\delta\epsilon_P$ between
the two peaks as a function of $\Phi^x_Q$.  We then fit those results
to the eigenspectrum of the Hamiltonian in
Eqn.~\ref{eqn:1qhamiltonian}, using $\Delta_Q$ and $\big|I^p_Q\big|$
as free parameters.

\begin{figure}
\includegraphics{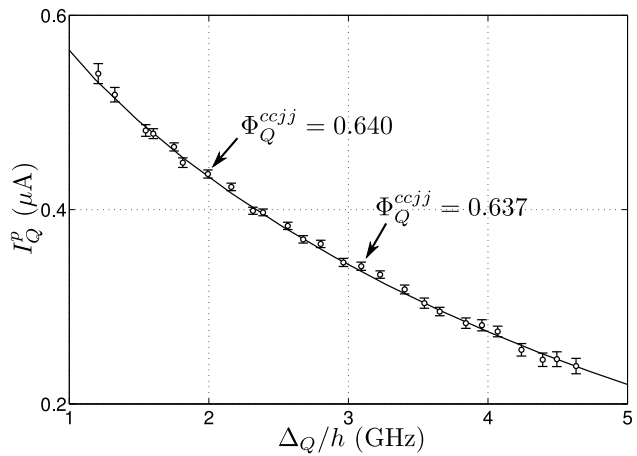}
\caption{\label{fig:ipanddelta}  $I^p_Q(\Phi^{ccjj}_Q)$ vs $\Delta_Q(\Phi^{ccjj}_Q)$
  from the $Q$ spectra for a range of settings of the control bias
  $\Phi^{ccjj}_Q$.  The data points from the fits in Fig.~\ref{fig:sample-1q-spectrum} (d) are labelled
  with arrows. The data also have horizontal error bars approximately
  the size of the symbols.  Results have been fit to the rf SQUID model
  from Ref.~\onlinecite{robustqubit} (solid line) with the rf SQUID capacitance $C$ as   
  the only free parameter.
}
\end{figure}

In order to crosscheck our QTS results, we have repeated the measurements and analysis that led to Fig.~\ref{fig:sample-1q-spectrum}(d) for several values of $\Phi^{ccjj}_Q$.  By doing so, we generated maps of the qubit parameters $\Delta_Q$ and $\left|I^p_Q\right|$ as a function of $\Phi^{ccjj}_Q$.  A plot of the relationship between $\left|I^p_Q\right|$ and $\Delta_Q$ is shown in Fig.~\ref{fig:ipanddelta}.  This curve is completely determined by the rf SQUID inductance $L$ and capacitance $C$ of qubit Q.  We have fit these results (solid line in Fig.~\ref{fig:ipanddelta}) to a physical rf SQUID Hamiltonian (Eqn. 4 in Ref.~\onlinecite{robustqubit}) taking $L =355.5\,$pH, as determined by independent measurements, and using $C$ as a free parameter.  The best fit returned $C =118\pm 2$ fF, which is a physically reasonable value, given Josephson junction sizes and qubit wiring geometry.  The single parameter fit models the data well, implying that QTS has correctly extracted the low energy spectrum of rf SQUID qubit $Q$.


QTS could, in principle, yield more than just the eigenspectrum of
system $S$.  In particular, there is significant information contained
in the spectral weight of the peaks in $\Gamma$.  For example, the
spectral lines inferred from Fig.~\ref{fig:sample-1q-spectrum}(b) are
less pronounced for the upper level at $\Phi^x_Q<0$ and for the lower
level at $\Phi^x_Q>0$.  This is due to the proportionality of the
initial transition rate $\Gamma$ to the small overlap of the initial
state of $Q$ ($\sim\mket{\sdownarrow}_Q$) with its final state ($\sim
\mket{\suparrow}_Q$).  Choosing $\epsilon_{\text{comp}}=+2J$, instead
of $-2J$, yields a system in which the states in \mdown\ exchange
roles, thus yielding a new initial state
$\mket{\suparrow}_Q\mket{\sdownarrow}_P$.  Repeating the QTS
experiment with this configuration should then swap the regions of
high and low peak visibility seen in
Fig.~\ref{fig:sample-1q-spectrum}(b).  Thus the spectral weight
contains information about the wavefunction of the probed system.

Further information could be gleaned from the lineshapes of
inter-manifold tunneling processes.  We chose to fit tunneling rate
peaks to Gaussians as we had anticipated that their lineshapes would
be dominated by the incoherent tunneling of the slow probe qubit $P$,
as in Ref.~\onlinecite{fluxnoisemrt}.  A detailed analysis of the
physical mechanisms that lead to particular lineshapes is currently
underway.  

We have demonstrated a low bandwidth method, termed qubit tunneling spectroscopy (QTS), by probing the energy spectrum of a first qubit by using a second probe qubit to split the two qubit system into two manifolds of qualitatively different states. Transitions between the two manifolds are monitored as a function of the energy bias of one of the manifolds.  Transition rate peaks correspond to the presence of eigenstates in the target manifold under study at that energy bias.  We validated this method by verifying that rf SQUID flux qubit energy spectra measured in this manner are consistent with an rf SQUID Hamiltonian.  QTS provided a direct measurement of the first qubit's energy splitting $\Delta_Q$ that was three orders of magnitude larger than the measurement bandwidth $\Delta_P$.

While the demonstration in this paper was limited to a single qubit, QTS is extensible to larger numbers of qubits and to other physical systems, provided one has good control and readout of the probe and a method of applying a compensation bias.  We anticipate that QTS will be a valuable tool for studying mesoscale systems.

The authors would like to acknowledge: F. Cioata, P. Spear for the
design and maintenance of electronics control systems; D. Bruce, P. deBuen,
M. Gullen, M. Hager, G. Lamont, L. Paulson, C. Petroff,
A. Tcaciuc for cryogenics and IO support; I. Perminov for software
design and support.

%


\end{document}